\documentclass[cameraready]{Interspeech}

\title{X-OPD: Cross-Modal On-Policy Distillation for Capability Alignment in Speech LLMs}

\author[affiliation={1}, equalcontribution]{Di}{Cao}
\author[affiliation={1,2}, equalcontribution, ]{Dongjie}{Fu}
\author[affiliation={1}]{Hai}{Yu}
\author[affiliation={1}]{Siqi}{Zheng}
\author[affiliation={1}, correspondingauthor]{Xu}{Tan}
\author[affiliation={2}, correspondingauthor]{Tao}{Jin}

\address{
    $^1$ Tencent Hunyuan $^2$ Zhejiang University
}

\email{tanxu2012@gmail.com, jint\_zju@zju.edu.cn}


\usepackage{comment}
\usepackage{amsmath}
\usepackage{booktabs}   
\usepackage{multirow}   
\usepackage{tabularx}   
\usepackage{caption}    
\usepackage{algorithm}
\usepackage{algorithmic}
\usepackage{amsmath}
\usepackage{graphicx} 
\usepackage{geometry}   
\usepackage{xcolor}

\begin{document}

\maketitle

\begin{abstract}
While the shift from cascaded dialogue systems to end-to-end (E2E) speech Large Language Models (LLMs) improves latency and paralinguistic modeling, E2E models often exhibit a significant performance degradation compared to their text-based counterparts. The standard Supervised Fine-Tuning (SFT) and Reinforcement Learning (RL) training methods fail to close this gap. To address this, we propose X-OPD, a novel Cross-Modal On-Policy Distillation framework designed to systematically align the capabilities of Speech LLMs to their text-based counterparts. X-OPD enables the Speech LLM to explore its own distribution via on-policy rollouts, where a text-based teacher model evaluates these trajectories and provides token-level feedback, effectively distilling teacher's capabilities into student's multi-modal representations. Extensive experiments across multiple benchmarks demonstrate that X-OPD significantly narrows the gap in complex tasks while preserving the model's inherent capabilities. 
\end{abstract}

\section{Introduction}

As multimodal LLMs become increasingly popular, speech interaction is transitioning from traditional cascaded architectures — typically comprising Automatic Speech Recognition (ASR), an LLM, and Text-to-Speech (TTS) — toward End-to-End (E2E) paradigms. By modeling directly within the continuous speech signal space, E2E models significantly reduce interaction latency and capture rich paralinguistic information, such as intonation, emotion, and environmental context, thereby offering a user experience that is far more expressive and akin to human-to-human interaction. This potential is exemplified by recent flagship systems — including GPT-4o \cite{hurst2024gpt}, Gemini 2.5 \cite{comanici2025gemini}, Qwen3-Omni \cite{xu2025qwen3} and Voxtral \cite{liu2025voxtral} — which have collectively redefined unified audio-text understanding.

However, despite the superior interaction fluency of E2E architectures, the significant performance gap remains a barrier to their widespread deployment. Empirical studies \cite{lu25c_interspeech, chen2025aha, fang2025s2sbench, zhang2025echox} indicate that these models frequently exhibit significant degradation in complex instruction following, logical reasoning, or knowledge-intensive queries compared to their text-based counterparts. Consequently, while cascaded systems may be slower, they remain the industrial selection due to the robust capabilities inherited from the underlying text LLM. This misalignment between modalities prevents speech models from fully leveraging the cognitive power of their textual foundations.

We attribute this degradation primarily to two factors: the scarcity of high-quality, paired speech-reasoning data, and the inherent misalignment between continuous acoustic representations and the discrete logical space of text LLMs. As a result, the high-quality Supervised 
Fine-Tuning (SFT) and Reinforcement Learning (RL) data in text LLM training cannot be fully transferred into the training of a Speech LLM, resulting in a gap in the standard SFT+RL training pipeline. Consequently, there is an urgent need for training paradigms that can align cross-modal capabilities without heavy reliance on static datasets.

Some methods try to bridge this gap by performing offline distillation from a cascaded system. However, these methods often struggle with distribution shifts — known as the exposure bias problem — where the model's generation trajectory during inference diverges from the training distribution. Furthermore, the accumulated errors in cascaded pipelines will also harm the performance. 


Addressing these challenges, we propose X-OPD, a novel Cross-Modal On-Policy Distillation framework designed to systematically bridge the intelligence gap between Speech LLMs and their text-based counterparts. We introduce a training strategy that utilizes transcribed text as an alignment bridge. In our framework, the student model performs autonomous rollouts across both speech and text modalities. Simultaneously, a more capable text-based teacher model generates a reference distribution based on the synchronized text input. By leveraging Kullback-Leibler (KL) divergence, denoted as $D_{\text{KL}}(P \parallel Q) $, for dynamic credit assignment, we precisely distill the teacher's logical knowledge into the student's multimodal representations. This approach offers three distinct advantages: it significantly enhances training efficiency; it eliminates the dependency on ground truth data, allowing for the utilization of open-source models where training data is undisclosed; and it minimizes the catastrophic forgetting of other acoustic capabilities. Through this framework, we aim to achieve a low-cost, high-efficiency alignment of cross-modal intelligence, paving the way for the next generation of smart, expressive spoken language agents.

In this work, we introduce X-OPD, a novel optimization strategy that achieves robust modality alignment while effectively preserving the pre-trained model's general proficiency. We demonstrate that X-OPD consistently outperforms various training paradigms and distillation benchmarks, significantly narrowing the performance gap.

\section{Related Work}
\subsection{Distillation in Speech LLMs}
Recent studies utilize distillation to bridge the performance gap in Speech LLMs. DeSTA2.5-Audio \cite{lu2025desta2} achieves cross-modal alignment by employing  backbone LLM to generate responses from audio captions, which then serve as ground truth targets for SFT. SALAD \cite{cuervo2025closing} employs a two-stage approach, integrating cross-modal distillation with active data selection to achieve high sample efficiency. Wang et al. \cite{wang2025cross} introduced a dual-channel distillation framework with logit-level supervision to transfer complex reasoning capabilities from text teachers. These methods are fundamentally off-policy, relying on static teacher trajectories or fixed targets rather than the model's own inference rollouts. This results in exposure bias, as the model cannot learn to correct its own reasoning path upon deviation.

Qwen3-Omni claimed to achieve state-of-the-art performance across multiple modalities \cite{xu2025qwen3}, without a cross-modality gap. However, when evaluated on mainstream audio understanding benchmarks, such as Big Bench Audio and Audio Multi-Challenge, performance drops are observed compared to the Qwen3 counterpart \cite{yang2025qwen3}. This result further suggests that standard SFT+RL training schemes are insufficient to close the gap.  

\subsection{On-Policy Distillation}
To address the exposure bias in off-policy distillation, some research has pivoted towards on-policy paradigms for generative models. GKD \cite{agarwal2024policy} introduces Generalized Knowledge Distillation, which addresses the distribution mismatch in auto-regressive models by training the student on its self-generated sequences using teacher feedback. MiniLLM \cite{gu2024minillm} utilizes an on-policy distillation strategy with a reverse KL divergence objective to mitigate exposure bias and improve generation calibration in smaller language models. In some recent works within the industry, Qwen3 \cite{yang2025qwen3,xu2025qwen3} introduces a strong-to-weak distillation framework to transfer reasoning capabilities from flagship teachers to smaller students with exceptional efficiency. Thinking Machines Lab \cite{lu2025onpolicydistillation} explores on-policy distillation as a method combining the dense rewards of distillation with the on-policy relevance of RL, achieving massive compute efficiency gains over traditional RL.

\begin{figure}[t] 
    \centering
    \includegraphics[width=\columnwidth]{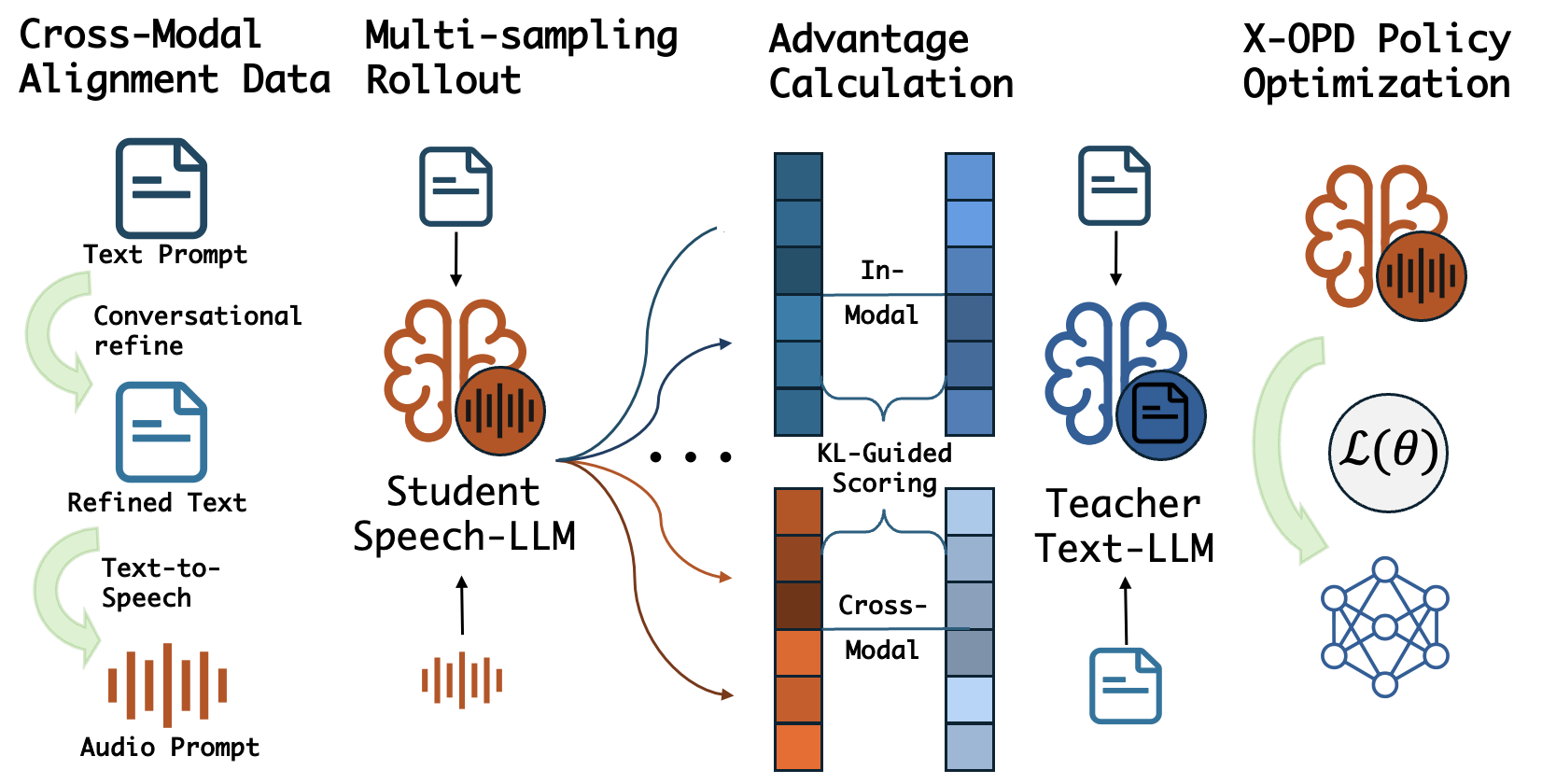} 
    \caption{Overview of the proposed Cross-Modal On-Policy Distillation (X-OPD) framework.}
    \label{fig:framework}
\end{figure}

\begin{table*}[t]
\centering
\caption{\textbf{Performance comparison across different model series and training paradigms.} We report performance scores across multiple benchmarks for both speech-input (S) and text-input (T) modalities. The \textbf{Avg. Drop (\%)} measures the aggregate performance gap relative to each series' original Base Model (for Gemini, its own text modality serves as the baseline).}
\label{tab:main_results_degradation}
\footnotesize
\setlength{\tabcolsep}{6pt} 

\begin{tabular*}{\textwidth}{@{\extracolsep{\fill}}l cc cc cc cc}
\toprule
\multirow{2}{*}{\textbf{Model Series}} & \multicolumn{2}{c}{\textbf{BIG Bench Audio}} & \multicolumn{2}{c}{\textbf{Audio Multi-Chall.}} & \multicolumn{2}{c}{\textbf{\hphantom{xx}Voice Bench\hphantom{xx}}} & \multicolumn{2}{c}{\textbf{Avg. Drop (\%)}} \\
\cmidrule(lr){2-3} \cmidrule(lr){4-5} \cmidrule(lr){6-7} \cmidrule(lr){8-9}
& $S$ & $T$ & $S$ & $T$ & $S$ & $T$ & $\Delta S$ & $\Delta T$ \\ 
\midrule

\textit{GPT-4o\footnotemark[1] (Base, Text-modality)} & -- & 90.27 & -- & 32.13 & -- & 92.61 & -- & -- \\ 
\quad GPT-4o-Audio\footnotemark[2] & 88.50 & -- & 24.57 & -- & 89.69 & -- & 9.55 & -- \\ 
\midrule

Gemini-2.5-Flash & 81.42 & 98.50 & 26.37 & 43.22 & 86.13 & 94.32 & 21.67 & -- \\ 
Gemini-3-Flash-Preview & 80.77 & 95.28 & 42.70 & 49.21 & 92.21 & 95.32 & 10.57 & -- \\ 
\midrule

\textit{Ministral-3-3B (Base, Text-modality)} & -- & 59.70 & -- & 22.67 & -- & 76.03 & -- & -- \\ 
\quad Voxtral-Mini-3B & 47.73 & 59.80 & 12.80 & 17.27 & 49.55 & 71.41 & 32.81 & 9.91 \\ 
\midrule

\textit{Qwen2.5-7B-Instruct (Base, Text-modality)} & -- & 74.07 & -- & 17.00 & -- & 81.06 & -- & -- \\ 
\quad Omni-7B-Instruct & 62.22 & 66.49 & 13.56 & 12.66 & 72.35 & 74.77 & 15.66 & 14.51 \\ 
\midrule

\textit{Qwen3-A3B-Instruct (Base, Text-modality)} & -- & 97.72 & -- & 29.31 & -- & 91.00 & -- & -- \\ 
\quad Omni-A3B-Instruct & 85.67 & 92.10 & 23.57 & 26.14 & 89.22 & 91.04 & 11.29 & 5.51 \\ 
\quad \quad + SFT & 87.52 & 85.30 & 15.04 & 19.25 & 82.66 & 86.06 & 22.76 & 17.49 \\ 
\quad \quad + Offline KD & 87.08 & 86.18 & 17.73 & 21.27 & 83.31 & 85.71 & 19.62 & 15.02 \\ 
\quad \quad + GKD (Forward KL) & 84.26 & 85.43 & 18.13 & 20.76 & 83.03 & 85.82 & 20.23 & 15.81 \\ 
\quad \quad \textbf{+ X-OPD (Ours)} & \textbf{93.41} & \textbf{96.85} & \textbf{28.14} & \textbf{28.66} & \textbf{89.29} & \textbf{91.18} & \textbf{3.43} & \textbf{0.97} \\ 

\bottomrule
\end{tabular*}
\end{table*}

\section{Cross-Modal On-Policy Distillation}
The overall framework of our proposed Cross-Modal On-Policy Distillation (X-OPD) is shown in Figure \ref{fig:framework}. X-OPD leverages on-policy sampling from the student model across both modalities, guided by token-level scoring from the text-based teacher. Optimized by policy gradients, X-OPD ensures Speech LLMs efficiently inherit the teacher's capabilities through direct cross-modal alignment.

\subsection{Cross-Modal Alignment Data}
To enable effective distillation, we define a parallel dataset $\mathcal{D} = \{ (S_i, T_i) \}$, consisting of paired speech and text prompts. The fundamental requirement for $\mathcal{D}$ is Semantic Invariance: the acoustic signal $S_i$ and the textual instruction $T_i$ must be strictly aligned in their logical intent. In practice, such alignment can be established by either synthesizing speech from textual instructions or transcribing existing audio prompts via speech recognition.

\subsection{Robust Multi-sampling Rollout}
Single-sample rollouts often suffer from inherent stochasticity, yielding excessive variance in gradient estimation that can destabilize the training process. To enhance optimization robustness, the policy independently samples $n$ candidate trajectories for each prompt. By marginalizing the gradients across these multiple paths, we achieve broader coverage of the policy space and significantly attenuate the high volatility characteristic of on-policy reinforcement learning updates.

\subsection{In-modal and Cross-modal Advantage Function}
To ensure the student model $\pi_{\theta}$ faithfully inherits the teacher's capabilities, we introduce a dual-advantage mechanism. This framework first employs an in-modal advantage to stabilize the student's foundational proficiency within the textual domain, providing a consistent reference for cross-modal alignment.

For a given trajectory $y$ sampled from the student policy $\pi_{\theta}$, let $y_t$ be the $t$-th token. The in-modal advantage $A_{im}(y_t)$ measures the log-probability discrepancy between the teacher $\pi_{\phi}$ and the student $\pi_{\theta}$ when both are conditioned on the text prompt $T$:

\begin{equation}
A_{im}(y_t) = \log \pi_{\phi}(y_t | T, y_{<t}) - \log \pi_{\theta}(y_t | T, y_{<t})
\end{equation}

Building on this, the cross-modal advantage $A_{cm}(y_t)$ is defined to bridge the gap between the teacher's textual logic and the student's speech-conditioned output:

\begin{equation}
A_{cm}(y_t) = \log \pi_{\phi}(y_t | T, y_{<t}) - \log \pi_{\theta}(y_t | S, y_{<t})
\end{equation}

Together, these terms provide a cross-modal reward signal, anchoring the student's alignment to the teacher's established proficiency.

\subsection{X-OPD Optimization Objective}
Integrating the aforementioned mechanisms, the optimization objective is formulated as a policy gradient loss comprising two components: the in-modal loss $\mathcal{L}_{im}$ and the cross-modal loss $\mathcal{L}_{cm}$. To estimate the gradient with reduced variance, the policy samples $m$ independent rollouts $y_j$ for each input through multi-sample rollout. The individual loss terms are defined as:

\begin{equation}
\mathcal{L}_{im}(\theta) = \mathbb{E}\left[ \frac{1}{m} \sum_{j=1}^{m} \frac{1}{|y_j|}\sum_{t=1}^{|y_j|} r_{j,t}(\theta) A_{im}(y_{j,t}) \right]
\end{equation}

\begin{equation}
\mathcal{L}_{cm}(\theta) = \mathbb{E}\left[ \frac{1}{m} \sum_{j=1}^{m} \frac{1}{|y_j|}\sum_{t=1}^{|y_j|} r_{j,t}(\theta) A_{cm}(y_{j,t}) \right]
\end{equation}

In these formulations, $y_j = \{y_{j,1}, \dots, y_{j,|y_j|}\}$ denotes the $j$-th trajectory sampled from the behavior policy. The term $r_{j,t}(\theta) = \frac{\pi_{\theta}(y_{j,t} | \cdot)}{\pi_{\text{old}}(y_{j,t} | \cdot)}$ represents the probability ratio between the current policy $\pi_{\theta}$ and the sampling policy $\pi_{\text{old}}$.

The final optimization objective for X-OPD is defined as the weighted sum of these two components:

\begin{equation}
\mathcal{L}(\theta) = \lambda \mathcal{L}_{im}(\theta) + (1 - \lambda) \mathcal{L}_{cm}(\theta)
\end{equation}

where $\lambda \in [0, 1]$ is a balancing hyperparameter.

\begin{table*}[t]
\centering
\caption{\textbf{Ablation Study of X-OPD.} The results are presented in the same format as Table \ref{tab:main_results_degradation}. The default X-OPD configuration is based on the Qwen3-Omni-A3B-Instruct, utilizing Qwen3-A3B-Instruct as the teacher with  $\lambda=0.5$. }
\label{tab:main_results_ablation}
\footnotesize
\setlength{\tabcolsep}{5pt} 

\begin{tabular*}{\textwidth}{@{\extracolsep{\fill}}l cc cc cc cc}
\toprule
\multirow{2}{*}{\textbf{Method}} & \multicolumn{2}{c}{\textbf{BIG Bench Audio}} & \multicolumn{2}{c}{\textbf{Audio Multi-Chall.}} & \multicolumn{2}{c}{\textbf{Voice Bench}} & \multicolumn{2}{c}{\textbf{Avg. Drop (\%)}} \\
\cmidrule(lr){2-3} \cmidrule(lr){4-5} \cmidrule(lr){6-7} \cmidrule(lr){8-9}
& $S$ & $T$ & $S$ & $T$ & $S$ & $T$ & $\Delta S$ & $\Delta T$ \\ 
\midrule

\textit{Reference (Text-only Models)} \\

Qwen3-A22B-Instruct & -- & \textbf{99.10} & -- & \textbf{35.73} & -- & \textbf{94.09} & -- & -- \\ 
Qwen3-A3B-Instruct & -- & 97.72 & -- & 29.31 & -- & 91.00 & -- & -- \\ 
\midrule

Qwen3-Omni-A3B-Instruct & 85.67 & 92.10 & 23.57 & 26.14 & 89.22 & 91.04 & 11.29 & 5.51 \\ 
X-OPD & \textbf{93.41} & \textbf{96.85} & \textbf{28.14} & \textbf{28.66} & 89.29 & 91.18 & \textbf{3.43} & \textbf{0.97} \\

\quad w/ Qwen3-A22B-Instruct & 91.91 & 95.29 & 26.62 & 27.86 & 89.60 & \textbf{91.67} & 5.55 & 2.23 \\ 

\quad $\lambda=1.0$ (Text-only OPD) & 91.76 & 96.01 & 25.44 & 27.62 & \textbf{89.70} & 91.21 & 6.91 & 2.43 \\ 
\quad $\lambda=0.0$ (Speech-only OPD) & 92.74 & 95.22 & 27.02 & 27.74 & 88.89 & 91.31 & 5.08 & 2.52 \\ 

\bottomrule
\end{tabular*}
\end{table*}

\section{Experiments}
\footnotetext[1]{Version: gpt-4o-2024-11-20}

\footnotetext[2]{Version: gpt-4o-audio-preview-2025-06-03}

\subsection{Training Data}
Our training set is derived from text-based prompts sampled from Tulu 3 \cite{lambert2025tulu} and NaturalReasoning \cite{yuan2025naturalreasoning}. First, we utilize Gemini-3-Flash-Preview to rewrite the original prompts into a more conversational and spoken-style format. These refined instructions are then synthesized into audio via CosyVoice3 \cite{du2025cosyvoice3, lyu2025build} at a 24kHz sampling rate. To ensure high acoustic fidelity, we employ SenseVoice \cite{an2024funaudiollm} for ASR back-translation, filtering out any samples with a Word Error Rate (WER) exceeding 5\%. The resulting dataset comprises two high-quality subsets: 10,934 pairs from Tulu 3 (136.7 hours) and 16,913 pairs from NaturalReasoning (95.5 hours), providing a robust parallel corpus of audio-based general instructions.

\subsection{Evaluation}
To provide a comprehensive assessment, we conduct evaluations on three representative speech benchmarks:

\begin{itemize}
    \item \textbf{BIG Bench Audio} evaluates high-level reasoning by adapting four logical tasks from BIG-bench Hard \cite{srivastava2022beyond, suzgun2023challenging} to the audio domain, provided by Artificial Analysis\footnote[3]{\url{https://artificialanalysis.ai/}}. It comprises 1,000 synthetic audio samples, with performance measured by accuracy.

    \item \textbf{Audio Multi-Challenge} \cite{gosai2025audio} tests natural multi-turn interaction capabilities including inference memory, instruction retention, self-coherence, and voice editing. It features 1,712 rubrics across 452 conversations. Performance is reported via the Average Pass Rate (APR).

    \item \textbf{VoiceBench} \cite{chen2024voicebench} evaluates general knowledge, instruction following, and safety through seven subsets (e.g., SQ-QA \cite{faisal2021sd}, IFEval \cite{zhou2023instruction}, and AdvBench \cite{zou2023universal}). Comprising 5,783 single-turn samples recorded under diverse acoustic conditions, the performance is reported as the mean score across all subsets.

\end{itemize}

Each benchmark provides paired speech and text instructions with ground-truth answers. While text-only models are limited to text inputs, speech-capable models are evaluated on both modalities separately. We follow official inference guidelines and employ Qwen3-235B-A22B-Instruct-2507 as the primary LLM-as-Judge. To mitigate stochasticity, we conduct three independent inference runs per benchmark, with each output evaluated three times. The final metrics represent the mean across these nine instances.

\subsection{Experimental Setup}
We employ Qwen3-Omni-A3B-Instruct as our base model due to its widespread adoption and robust performance within the research community. We implement X-OPD as an RL-style policy optimization using the verl framework \cite{sheng2025hybridflow}. The optimization follows a pure on-policy sampling strategy, where all responses are generated by the current policy. We perform full-parameter training on the language model backbone, while the audio tower and modal adapter are frozen throughout the optimization process. Training is conducted using the Adam optimizer with a learning rate of $2 \times 10^{-6}$ and a batch size of 256. To balance exploration and stability, we set the number of rollouts to $n=4$ per prompt.

For comparison, we establish three baselines using the ms-swift framework \cite{zhao2025swift}:  (1) Standard SFT, performed on the original ground-truth responses; (2) Offline Knowledge Distillation (KD), where the model is fine-tuned on responses generated by its text-only counterpart, Qwen3-A3B-Instruct; and (3) Generalized Knowledge Distillation (GKD) \cite{agarwal2024policy}, which optimizes the student policy based on the forward KL divergence relative to the Qwen3-A3B-Instruct teacher. All baseline models are trained for 1 epoch with hyperparameters consistent with the X-OPD setup.

\subsection{Results}
Table \ref{tab:main_results_degradation} presents a comparative analysis of various models, including GPT-4o, Gemini 2.5/3, Voxtral-Mini, and the Qwen2.5/3-Omni series. A consistent trend is observed: the majority of evaluated Speech LLMs suffer from performance degradation relative to their text-based counterparts. This discrepancy persists even under text-only evaluation, highlighting the fundamental challenges of modality alignment in current Speech LLM development, which hinders the preservation of the base model's general capabilities during speech-text cross-modal learning.

Notably, experimental results reveal that standard optimization strategies — namely SFT, Offline KD and GKD — counter-intuitively exacerbate performance degradation, despite being trained on the identical dataset as X-OPD. This suggests that naive alignment attempts may conflict with the base model's internal priors rather than harmonizing the two modalities. In contrast, X-OPD demonstrates superior efficacy in mitigating this ``alignment-tax," reducing the average performance drop for Qwen3-Omni-A3B-Instruct from 11.29\% to 3.43\% for Speech (S) and from 5.51\% to a mere 0.97\% for Text (T). This recovery is most pronounced in challenging tasks like BIG Bench Audio and Audio Multi-Challenge, whereas on Voice Bench, X-OPD maintains parity with the base model, effectively reaching the performance ceiling. These findings underscore X-OPD's unique capability to achieve seamless modality alignment while fully preserving the base model's original general capabilities.

\subsection{Ablation Study}
Table ~\ref{tab:main_results_ablation} summarizes our ablation study on Qwen3-Omni-A3B-Instruct, evaluating the impact of teacher capacity and the balancing hyperparameter $\lambda$. Regarding teacher capacity, although Qwen3-A22B-Instruct possesses stronger raw capabilities, using the A3B version yielded superior performance. 
This suggests that effective alignment relies less on maximizing teacher scale and more on matching capacities to mitigate the ``knowledge gap" ~\cite{li2025small}, thereby providing more aligned and accessible trajectories ~\cite{xu2024stronger}.
In terms of objective balancing, results for $\lambda$ reveal a reciprocal reinforcement between modalities: textual distillation ($\lambda=1.0$) benefits speech performance, while speech-focused distillation ($\lambda=0.0$) conversely enhances text-only scores. This cross-modal synergy demonstrates that the two objectives complement rather than compromise general capabilities. Consequently, the balanced setting ($\lambda=0.5$) yields superior overall results, validating that X-OPD's dual-objective approach is essential for maximizing this synergy and achieving peak performance.

\subsection{Analysis of Catastrophic Forgetting}
To investigate whether the model retains its foundational capabilities, we evaluate catastrophic forgetting using the MMAR benchmark \cite{ma2025mmar}. Assessing reasoning across 1,000 curated samples of sound, music, and speech, MMAR serves as an ideal proxy for measuring the retention of pre-trained knowledge.

\begin{table}[h]
\centering
\caption{Analysis of catastrophic forgetting on the MMAR benchmark. Total accuracy ($\uparrow$) is reported.}
\label{tab:mmar}
\small
\begin{tabular}{lcc}
\toprule
\textbf{Method} & \textbf{MMAR (\%)} & \textbf{Drop} \\ \midrule
Original Model & 71.3 & - \\ \midrule
SFT            & 60.3 & 11.0 \\
Offline KD             & 59.9 & 11.4 \\
GKD (Forward KL)            & 60.0 & 11.3 \\ \midrule
X-OPD ($\lambda=0$)   & 69.8 & 1.5 \\
X-OPD ($\lambda=0.5$) & 69.3 & 2.0 \\
X-OPD ($\lambda=1$)   & \textbf{70.7} & \textbf{0.6} \\ \bottomrule
\end{tabular}
\end{table}

As illustrated in Table~\ref{tab:mmar}, traditional training methods — including SFT, Offline KD, and GKD — suffer from severe performance degradation, with accuracies plummeting from 71.3\% to 59.9\%. In contrast, all X-OPD variants demonstrate remarkable robustness, maintaining near-lossless scores (all $>69\%$). Notably, the text-only distillation setting ($\lambda=1$) yields the highest retention at 70.7\%, confirming that in-modal optimization imposes the least interference on pre-trained audio features. Even when incorporating cross-modal objectives ($\lambda=0$ or $0.5$), X-OPD consistently maintains a marginal decline compared to the baseline methods. This demonstrates that X-OPD's RL-style paradigm effectively regularizes the model's behavior, successfully balancing cross-modal alignment with the preservation of its original general capabilities.

\section{Conclusion}
In this work, we propose X-OPD, a novel Cross-Modal On-Policy Distillation framework that effectively aligns Speech LLMs with their text-based counterparts. Our experiments show that X-OPD significantly narrows the performance gap between speech and text modalities while effectively mitigating catastrophic forgetting. Notably, X-OPD exhibits remarkable sample efficiency, achieving superior alignment and capability preservation with a modest dataset of only 27k samples. This establishes a robust, data-efficient, and annotation-free pathway for foundational alignment in multimodal agents. 

\section{Acknowledgments}

\ifcameraready
     The Interspeech 2026 organizers
\else
     The authors
\fi
would like to thank ISCA and the organizing committees of past Interspeech conferences for their help and for kindly providing the previous version of this template.

\section{Generative AI Use Disclosure}

We confirm that all authors are fully responsible and accountable for the work and content presented in this paper. Gemini 3 was employed exclusively for linguistic refinement and to enhance the clarity and flow of the manuscript. The authors emphasize that the AI was not used to generate any significant portion of the research.

\bibliographystyle{IEEEtran}
\bibliography{mybib}

\end{document}